\documentclass[12pt]{iopart}
\usepackage{graphicx}

\begin{document}
\title[Magnetic field effect in Pb-$p$-Hg$_{1-x}$Cd$_x$Te Schottky barriers]{Strong magnetic field effect on above-barrier transport
in Pb-$p$-Hg$_{1-x}$Cd$_x$Te Schottky barriers}

\author{V F Radantsev$^1$ and V V Zavyalov$^2$}

\address{$^1$ Institute of Natural Sciences, Ural Federal University, Ekaterinburg 620000, Russia}
\address{$^2$ Department of Physics,
Utah State University, 1600 Old Main Hill Logan, Utah 84322-4415,
USA}

\ead{victor.radantsev@usu.ru}

\begin{abstract}
It is usually supposed that the above-barrier current in Schottky
barriers on p-type semiconductor is controlled by the heavy holes.
However, in real structures, there is an additional potential
barrier caused by a oxide layer at interface. For typical values of
thickness and height of a barrier its tunnel transparency for light
holes can be higher by three order of magnitude than that for heavy
holes and one can expect that the current is manly a contribution of
light holes. To clear up this problem the investigation of transport
in a magnetic field is used as a key experiment in this work. The
pronounced magnetic field effect for heavy holes in investigated
Pb-$p$-Hg$_{1-x}$Cd$_x$Te Schottky barriers is expected only at
magnetic fields $B>10$ T. At the same time experimentally more than
twofold decrease in saturation current is observed even at
$B\sim0.5$ T at any orientation of magnetic field. The studies
performed for Hg$_{1-x}$Cd$_x$Te with different Kane's gap and at
different temperatures show that the magnitude of magnetic field
effect is uniquely determined by the ratio of light hole cyclotron
energy to a thermal energy $\theta=\hbar\omega_{clh}/kT$. However
the magnitude of effect exceeds considerably the prediction of the
simple theory. The reason of discrepancy remains a mystery.
\end{abstract}

 \submitto{\SST}
 \maketitle

\section {Introduction}

It is conventional to suppose that the heavy holes give the dominant
contribution to a above-barrier carrier transport in Schottky
barriers (SB) based on $p$-type semiconductors.  The above-barrier
current in ideal SB associated with the $j$th carrier type is
described by the expression \cite{Sze}
\begin{equation}\label{Ideal}
    I_j =I_{0j}\left[\exp{\left(\frac{eV_b}{kT}\right)}-1\right],
\end{equation}
where $e$ is the electronic charge, $T$ the absolute temperature,
$k$ Boltzmann constant, $V_b$ the voltage drop across barrier (the
voltage for forward bias is chosen as positive) and $I_{0j}$ is the
saturation current expressed by
 \numparts
 \begin{eqnarray}
    I_{0j} =
    Sep_j\overline{v}_j\exp{\left(\frac{-\varphi_{d0}}{kT}\right)}=SA^{*}_jT^2\exp{\left(\frac{-\varphi_{B}}{kT}\right)}\label{Ideal2a}
 \end{eqnarray}
 or
 \begin{eqnarray}
 I_{0j} &=
    \nonumber Sep_jv_{dj}\exp{\left(\frac{-\varphi_{d0}}{kT}\right)}=\\
    &=SA^{*}_jT^{3/2}\left(\frac{4\pi m(\varphi_{d0}-eV_b)}{k\varepsilon\varepsilon_0}\right)^{1/2}
    \exp{\left(\frac{-\varphi_{B}}{kT}\right)}\label{Ideal2b}
 \end{eqnarray}
 \endnumparts
within the framework of diode (thermionic emission) or diffusion
theory respectively. In equations (\ref{Ideal2a}) and
(\ref{Ideal2b}), $S$ is the diode area, $p_j$ the concentration of
the $j$th carrier type, $\overline{v}_j=\sqrt{kT/2\pi m^{*}_j}$
average thermal velocity, $m^{*}_j$ the effective mass,
$A_j^{*}=ek^2m^{*}_j/2\pi^2\hbar^3$ effective Richardson constant,
$\hbar$ Planck constant, $\varphi_{d0}=\varphi_{B}-\eta$ diffusion
potential (band bending) at zero bias, $\varphi_{B}$ Schottky
barrier height, $\eta$ energy difference between the bulk Fermi
level and valance band edge, $v_{dj}=\mu_jE_m$ drift velocity,
$\mu_j$ mobility, $E_{m}$ maximum electric field in the space-charge
region, $\varepsilon$ and $\varepsilon_0$ the permittivities of the
semiconductor and free space.

The "classical" argument for neglecting the contribution of the
light holes in thermionic current in SB is based on the fact that
the ratio between the partial heavy holes (hh) and light holes (lh)
currents
 \begin{equation}\label{m/m}
 \frac{I_{hh}}{I_{lh}}=\frac{p_{hh}\overline{v}_{hh}}{p_{lh}\overline{v}_{lh}}=\frac{m^{*}_{hh}}{m^{*}_{lh}}\left[1+\frac{15}{4}\frac{kT}{E_{g}}\left(
  1+\frac{kT}{E_{g}}\right)
 \right]^{-1}
 \end{equation}
is proportional to the ratio of carrier effective masses (the factor
in brackets in the equation (\ref{m/m}) arises because of
nonparabolicity of light holes band, $E_g$ is Kane's energy gap). In
the case of Hg$_{1-x}$Cd$_x$Te narrow gap semiconductor the light
holes must contribute according to this assumption less than 5$\%$
to total current even for $x=0.3$. However, the above argumentation
is not justified for the real SB as we are to take into
consideration a thin oxide layer existing as a rule at the
interface. In the presence of the dielectric layer the current
density is reduced by the value of the quantum-mechanical
penetration coefficient P which in the case of rectangular potential
barrier shape is given in WBK approximation by
\begin{equation}\label{Penter}
    P_j=\exp\left(-\frac{2\sqrt{2}}{\hbar}\delta
    \sqrt{m^{*}_{tj}\varphi_{ox}}\right) ,
\end{equation}
where $m^{*}_{tj}\approx (1/m_0 +1/m^{*}_j)^{-1}$ is the tunnelling
(reduced) mass \cite{Tun_mass} for the $j$th carrier type, $m_0$ the
rest mass of the electron, $\delta$ and $\varphi_{ox}$ are the
thickness and barrier height of the interfacial insulator layer
respectively. For typical thickness $\delta=1$ nm and reasonable
value of barrier height $\varphi_{ox}=1$ eV the equation
(\ref{Penter}) gives $P_{hh}\sim 7\times10^{-4}$ and $P_{lh}\sim
0.36$ for heavy ($m^{*}_{thh}\sim 0.5m_0$) and light
($m^{*}_{tlh}\sim 0.01m_0$) holes respectively. So an oxide layer is
transparent for light carriers and practically impenetrable for
heavy carriers. If the presence of potential barrier of the
interfacial insulator is taken into account the ratio between the
heavy and light hole current is then given by
 \begin{eqnarray}\label{I/I}
 \frac{I_{hh}}{I_{lh}}=&\frac{p_{hh}\overline{v}_{hh}P_{hh}}{p_{lh}\overline{v}_{lh}P_{lh}}
 \approx\frac{m^{*}_{hh}}{m^{*}_{lh}}\left[1+\frac{15}{4}\frac{kT}{E_{g}}\left(
 1+\frac{kT}{E_{g}}\right) \right]^{-1} \nonumber\\
 &\times\exp\left[-\frac{2\sqrt{2}}{\hbar}\delta
   \varphi_{ox}\sqrt{m^{*}_{hh}}\left(1-\sqrt{m^{*}_{lh}/m^{*}_{hh}}\right)\right].
\end{eqnarray}
For the $\delta$, $\varphi_{ox}$, $m^{*}_{lh}$ and $m^{*}_{hh}$
values above chosen a value of 0.07 is obtained for ratio
$I_{hh}/I_{lh}$ ($T=100$ K). Thus for the real SB as compare with
ideal SB one can expect an inverse relation between the currents
arising from the light and heavy holes $I_{lh}\gg I_{hh}$. Generally
speaking, there are some indirect suggestions supporting this guess.
The analysis of literature date reveals the extremely small
experimental values of Richardson constant $A^{*}$ (by factor
$\sim10^2$) in SB Pb, Au, Ti, Al, Cr-$p$-HgCdTe \cite{Polla, Bahir,
RadFTP} in comparison with its theoretical value within an ordinary
assumption that the heavy holes are the majority current carriers in
SB on the base of $p$-semiconductor. However, in real SB the large
discrepancy between experimentally observed $A^{*}$ and the
theoretical value can be attributed to inferior diode quality. Even
for SB exhibiting low ideality factors, the extracted Richardson
constant can vary over orders of magnitude.

In order to examine unambiguously whether the light carriers
dominate in above-barrier transport one can use the magnetic field
effect on the current-voltage ($I-V$) characteristics. Two
quantities in the expression for current (\ref{Ideal}) can be
affected by a magnetic field: (i) Schottky barrier height
$\varphi_{B}$ because of an increase in effective energy gap
$E_g(B)=E_g(0)+\Delta E_g(B)$, where $\Delta E_g(B)$ is of the order
of a cyclotron energy $\hbar \omega_c=\hbar eB/m^{*}$ and (ii) the
mobility $\mu(B)=\mu(0)\left(1+\mu(0)^2B^2\right)^{-1}$ (in the case
of diffusion transport only). As evident from equations
\ref{Ideal2a} and \ref{Ideal2b}, the values of effect are determined
by the parameter $\vartheta =\hbar \omega_c/kT$ and $\mu(0)B$ in
first and second case respectively. Due to the small effective mass
and high mobility of light holes the magnetic fields should suppress
the light hole component of the current while the that of the heavy
hole is hardly affected. Therefore magnetic field application
enables a precise experimental separation between the two
components.

The influence of a magnetic field on the transport in Schottky
barriers was investigated in work \cite{RadSST} at low temperatures,
when the tunnel currents dominate. However we have not found any
works studying the magnetic field effect on the above-barrier
transport in SB. Most likely, it is caused by extremely small
magnitude of effect expected in Schottky barriers on the base both
of wide-gap and narrow-gap semiconductors, because in the latter
case the rectifying contacts are implemented only for the materials
of a $p$-type.

The results of the first study of the DC transport in
Pb-$p$-Hg$_{1-x}$Cd$_x$Te Schottky barrier in magnetic field at high
temperatures where the above-barrier transport predominates are
reported in this paper. As the light holes effective mass being a
critical parameter of a problem under consideration is alloy
composition dependent the samples with the different $x$ were
investigated.

\section {Details of experiment}
The used semiconductor substrates were $p$-type vacancy doped bulk
single crystals. Polished plates were etched in a 5\% solution of
Br$_2$ in methanol for 5-10 s with subsequent deposition of a think
insulating film providing mounting areas for the electrode of the
Schottky diodes. Directly before the evaporation of the metal, the
structures were subjected to a brief (1-2 s) etching in a solution
of the same composition as before (omission of one of these
operations resulted in a drastic deterioration of the Schottky
barrier characteristics). Typical SB areas were
$S=10^{-4}-2\times10^{-3}$ cm$^2$. Measurements were carried out on
the best-quality structures with the lowest values of ideality
factor and the smallest leakage current. The Schottky diodes with
such proprieties and a high reproducibility of the characteristics
were obtained only in the case of Pb-$p$-HgCdTe structures. The high
quality SB with Pb is evidently due to high values of the
tetrahedral radius  and of the saturation vapor pressure of Pb as
compared with other metals. These properties should minimize the
structural damage during evaporation and the diffusion of metal into
HgCdTe as its surface is known to be easily damaged because of the
weakness of the chemical bonds.

The alloy composition, energy gap and doping level of investigated
samples are listed in Table \ref{tabl}. The heavy holes mobility are
$\mu_{hh}=450, 400$ and $420$ (in cm$^2$ V$^{-1}$ s$^{-1}$) for
samples PK2, C37 and C36 correspondingly. The values for
$N_A-N_D\approx p_{hh}$ and $\mu_{hh}$ were determined from the Hall
measurements at 77 K. The sample temperature was monitored by using
a copper-constant thermocouple and temperature controller with
sensitivity better than 0.2 K.

\begin{table}
\caption{\label{tabl} Sample proprieties and basic Schottky barrier
characteristics. A effective Richardson constant $A_0^{*}$ and
Schottky barrier height $\varphi_{B}$ correspond to low temperature
portion of Richardson plot (see section \ref{Rich}).}
\begin{indented}
\lineup
\item[]
\begin{tabular}{@{}*{9}{l}}
 \br
 &$x$& $E_{g}$$^{\rm a}$ & $N_{A}-N_{D}$ & $T_{0}$ & $\varphi _{B}$$^{\rm a}$& $\beta $ & $A_0^{*}$ & $\Delta \varphi _{im}$$^{\rm b}$\cr
 sample && (meV) & (cm$^{-3}$) & (K) & (meV) &  & A$\cdot$m$^{-2}$K$^{-2}$ &(meV)\cr\mr
 PK2 &0.221 & \094 & 4.0$\times $10$^{15}$ & \080 & \075 & 1.15 &150 & 5.0 \cr
 C37 &0.290& 210 & 1.5$\times $10$^{15}$ & 110 & 160 & 1.05 & \060 & 5.3 \cr
 C36 &0.289& 208 & 6.0$\times $10$^{15}$ & 100 & 165 & 1.05 &\065  & 8.2 \cr
 \br
\end{tabular}
\item[]$^{\rm a}$ At $T=0$ K.
\item[]$^{\rm b}$ At zero bias and $T=140$ K.
\end{indented}
\end{table}

\section {Current-voltage characteristics in magnetic field}
\begin{figure}[t]
\begin{center}\leavevmode
\includegraphics[scale=1.3]{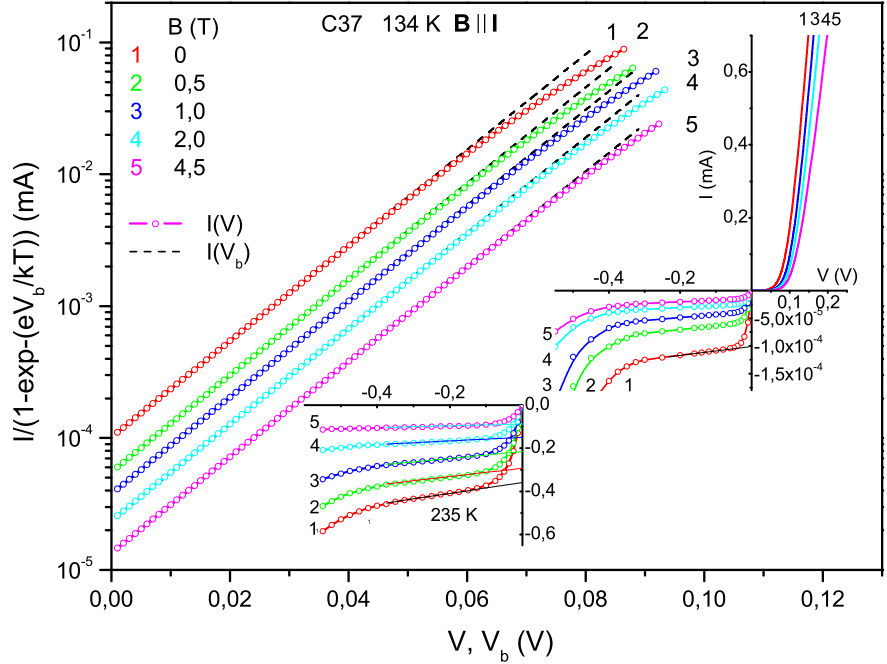}
\end{center}
\caption{Logarithmic plot of $I/\left[1-\exp\left(eV/kT\right)
\right]$ versus forward bias V and voltage drop across barrier
$V_b=V-IR_s$ ($R_s$ is the series resistance) at various magnetic
fields for sample C37. In the insets the experimental $I-V$
characteristics are shown.} \label{fig1}
\end{figure}
The typical current-voltage characteristics for sample C37 at high
temperatures $T>T_0$ at which the above-barrier transport dominates
(the typical $T_0$ values are listed in Table \ref{tabl}) are
presented in figure \ref{fig1} versus magnetic fields as a
parameter. In a nonideal SB, the $I-V$ characteristic deviates from
the simple expression (\ref{Ideal}). The leakages, barrier height
dependence on voltage, recombination in the space charge region,
tunnelling across the potential barrier, carrier trapping by
interface states modify the nature of the carrier transport. In
addition, part of the applied voltage drops across a series
resistance. An ideality factor, $\beta>1$, is commonly used to model
phenomenologically the effect of all the nonidealities ($\beta=1$
for pure thermionic emission) \cite{Sze,Rhoder}
\begin{equation}\label{nonIdeal}
    I = I_s\exp\left[\frac{e\left(V-IR_s\right)}{\beta kT}\right]\left[1 - \exp\left[\frac{e\left(V-IR_s\right)}{kT}\right]\right].
\end{equation}
As evidence in figure \ref{fig1} the forward branch of $I-V$
characteristics in studied SB is well described by the equation
(\ref{nonIdeal}) in the absence of a magnetic field as well as in
non-zero magnetic field. The ideality factor, extracted from the
linear part of the $\ln\left\{I/\left[1 -
\exp\left(eV_b/kT\right)\right]\right\}-V$ curves at $T>T_0$, is
practically temperature independent (see figure \ref{fig6}a) and
does not changed in magnetic field (at $T<T_0$  the tunnel currents
prevail and the value $\beta kT=E_0$ is independent of temperature
thus $\beta \propto 1/T$). The effect of magnetic field on forward
branch of $I-V$ characteristics is reduced to a decrease of
saturation current $I_s$ (the intercept points with ordinate axis in
figure \ref{fig1}).

At the same time as it can be seen in figure \ref{fig1} the reverse
current rises slowly (near-linearly at low bias) with the applied
bias. Nevertheless, though a reverse current does not show effect of
saturation, the value of $I$ corresponding to the intercept at the
ordinate of linear portion of revers branch of experimental $I-V$
characteristic is in close agreement with the saturation current
determined from a forward branch. The bias dependence of a reverse
current can be satisfactory explained if  the barrier height
lowering due to the image force  is taken into account \cite{Rhoder}
 \begin{eqnarray}\label{U-image}
 \Delta\varphi_{im}=\left[\frac{e^2 N_a}{8\pi^2 (\varepsilon
 \varepsilon_0)^3}\left(\varphi_B-\eta-eV-kT\right)\right]^{1/4},
\end{eqnarray}
In contrast to "classical" semiconductors in narrow-gap
Hg$_{1-x}$Cd$_x$Te (at $x<0.5$) the gap $E_g$ and consequently the
barrier height $\varphi_B$ increase with temperature \cite{Eg(T)}
 \begin{eqnarray}\label{Eg(xT)}
 E_g(x,T)=&-0.302+1.93x-0.810x^2+0.832x^3+5.35\cdot 10^{-4}\times\nonumber\\
 &\times(1-2x)\left[\left(T^3-1822\right)/\left(T^2+255.2\right)\right]
 \end{eqnarray}
(for $E_g$ in eV and $T$ in K). Assuming that the barrier height is
varied with temperature as the band gap such that Fermi level at
interface is a fixed percentage above the valence band edge, the
temperature dependence of $\varphi_B$ can be described as
\begin{equation}\label{Fi(T)}
    \varphi_B(T)=\varphi_B(0)+\gamma \alpha kT,
\end{equation}
where the temperature coefficient of energy gap $\gamma$ is
determined by relation (\ref{Eg(xT)}) $\gamma=\rmd E_g/\rmd
kT\approx 6.21(1-2x)$ and $\alpha$ is the ratio $\varphi_0/E_g$ at
zero temperature (see Table \ref{tabl}). The potential barrier
increase with temperature is clearly exhibited in the anomalous
temperature dependence of the reverse tunnelling current, which
dominates at $T<T_0$. The interband and trap-assisted tunnelling
currents in temperature range from $T=20$ K to $T=80$ K decrease by
a factor 10 and more. The dependence (\ref{Fi(T)}) agrees well with
the magnetic fields dependence of the tunnelling currents
investigated in similar Pb-$p$-HgCdTe SB in \cite{RadSST}. The
$\Delta\varphi_{im}$ values calculated for zero bias and $T=140$ K
are presented in table \ref{tabl}. The calculated Fermi energies
$\eta$ were obtained from the neutrality equation with allowance for
nonparabolicity in three-band Kane approximation.

\begin{figure}[t]
\begin{center}\leavevmode
\includegraphics[scale=1.3]{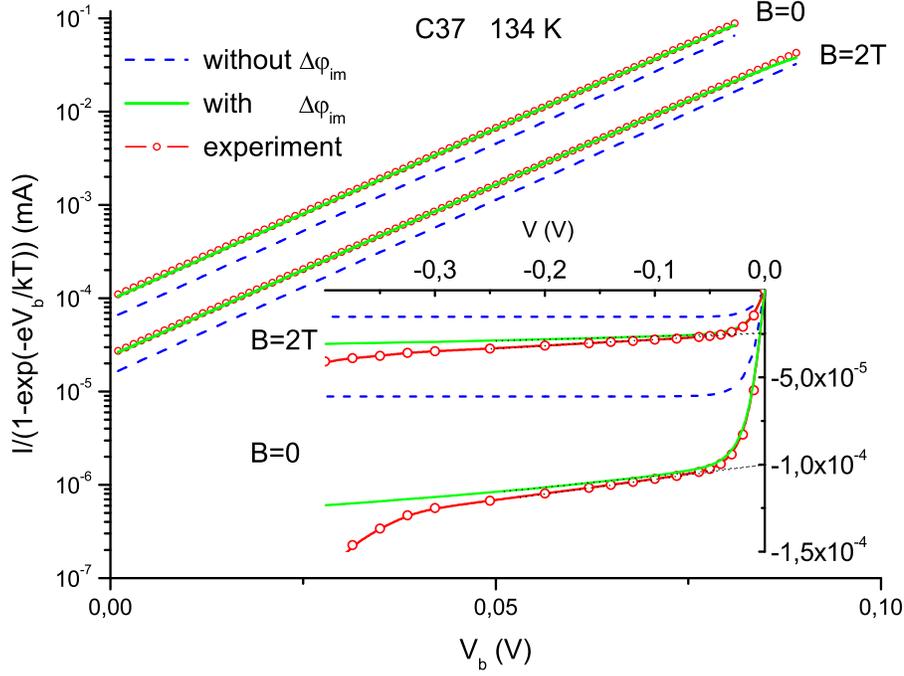}
\end{center}
\caption{The forward and revers (insert) voltage-current
characteristics with and without allowance for the barrier height
lowering due to the image force.} \label{fig2}
\end{figure}

The $I-V$ characteristics calculated with allowance for the image
force
\begin{equation}\label{Js_imag}
 \fl   I = I_0\exp\left[\frac{\Delta
    \varphi_{im}(V)}{kT}\right]\left[\exp{\left(\frac{eV}{kT}\right)}-1\right]=I_s(V)\left[\exp{\left(\frac{eV}{kT}\right)}-1\right]
\end{equation}
are compared with those measured experimentally in figure \ref{fig2}
for the same sample, temperature and magnetic fields as in figure
\ref{fig1}. It seen that at low bias the calculated reverse current
is indeed linearly bias dependent. In calculations the input
saturation current $I_0$ was chosen so that the intercepts at the
ordinate of linear portions of the output (calculated with allowance
for the image force) and the experimental reverse characteristics
$I_s$ coincided. Simultaneously the same accordance occurs for
forward branches also (see figure \ref{fig2}). The barrier height
lowering causes the slope of forward $\ln I-V$ characteristic to
increase by 4-6 $\%$ in accordance with experimental value of
ideality factor $\beta=1.04-1.05$. The modified (by barrier
lowering) saturation current $I_s$ is higher by factor $\approx$ 1.6
than the input (for ideal SB) saturation current $I_0$ . The
$I_s/I_0$ ratio decreases only slightly with increasing temperature.
For investigated temperature range $I_s/I_0$ ranges from 1.6 to 1.3
for samples PK2 and C37 and from 2 to 1.6 for more strongly doped
sample C36. The near-constant slope of the revers characteristics at
low bias can formally be considered as the effective leakage
conductance. Its value has temperature dependence close to the one
for saturation current, that agrees satisfactorily with experiment.
Because the function $\exp (\Delta \varphi_{im}/kT)$ appears in the
expression for current as magnetic fields independent multiplier,
$I_s$ varies with magnetic field exactly as does $I_0$ (figure
\ref{fig2}).

Though the barrier increase with temperature, diffusion potential
$\varphi_{d0}$ falls because of increasing of a Fermi energy $\eta$.
On the other hand, a saturation current exponentially increases with
temperature. This leads to an increase in a voltage drop across a
series resistance (in spite of the fact that the resistance
decreases with $T$ at high temperatures). Note, that the
contribution of the voltage dropped across the series resistance to
$I-V$ experimental plot can be reduced by the order of magnitude if
three-terminal method of measurement is applied (ohmic contacts on
the opposite sides of substrate wafer are used as a current contact
and potential probe respectively). The higher the diffusion
potential and lower the series resistance, the greater the range
over which the $\ln I-V$ curve yields at straight line. As range of
the exponential dependence of the forward $I-V$ plots shrinks with
temperature, the accuracy of the determination of saturation current
and $\beta$ from forward branch at high temperatures becomes poor.
In this case the relation $\ln((I-VG_l)/I_s+1)=e(V-IR_s)/\beta kT$
was fitted to experimental data in range of bias
$-2kT<eV_b<\varphi_{d0}$ using $I_s$ and $\beta$ as adjusting
parameters. The values $R_s$ and $G_l$ were determined from forward
and revers $I-V$ characteristics in sufficiently high bias regions.
The $I_s$ values obtained in such way do not contradict those
determined from the reverse  branches of $I-V$ characteristics.

\section {Magnetic field dependence of saturation current\label{IB}}

\begin{figure}[t]
\begin{center}\leavevmode
\includegraphics[scale=1.7]{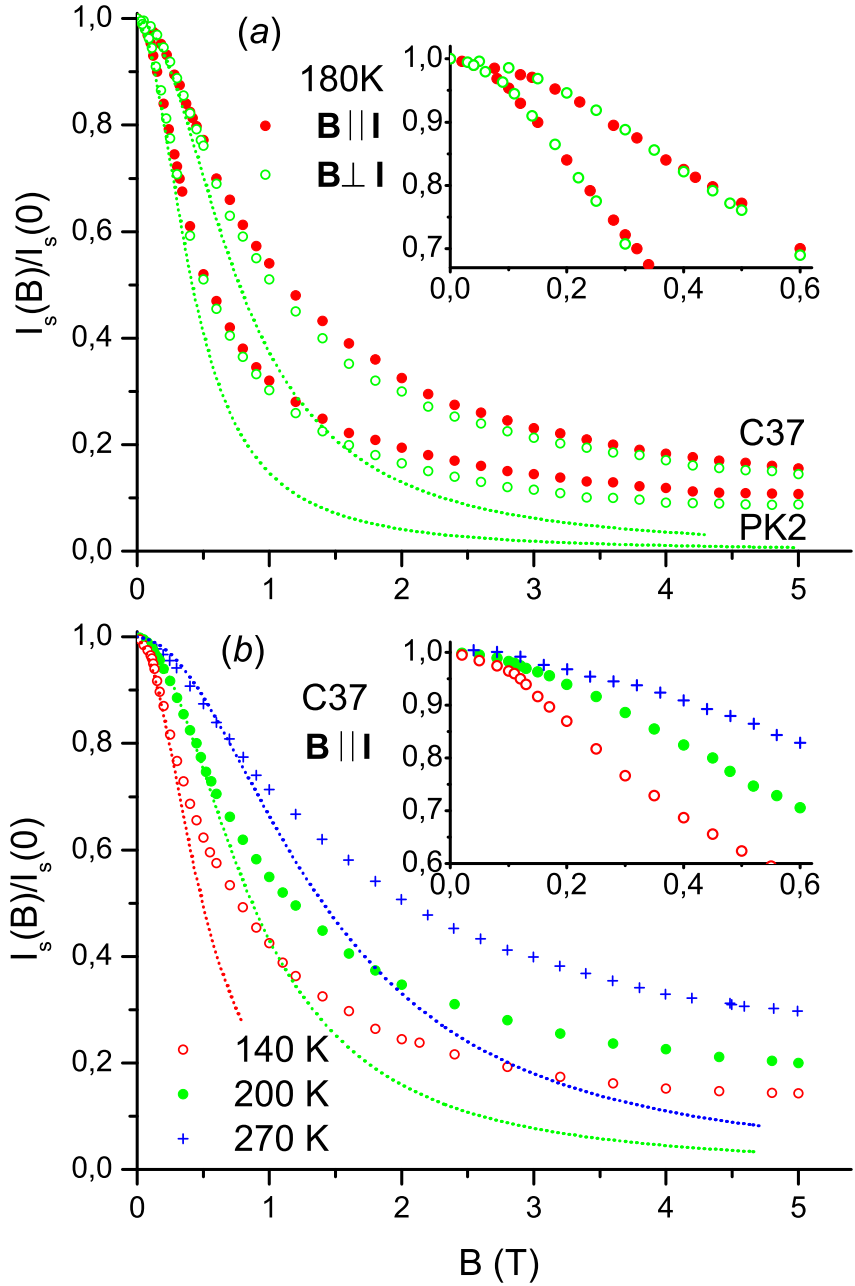}
\end{center}
\caption{Typical magnetic field dependence of the normalized
saturation current for samples with different light holes effective
mass at different temperatures. The dotted lines correspond to a
relation (\ref{Js_mobil}).} \label{fig3}
\end{figure}

To find out the character of $I_s(B)$ dependencies the current at
fixed forward and (or, at high temperatures) reverse bias were
recorded as a function of a magnetic field. The biases were chosen
so that they corresponded to the beginning of linear portions of
$I-V$ and $\ln I-V$ characteristics for reverse and forward biases
respectively. In the first case the correction for change of a slope
of $I-V$ plot in magnetic field was taken into account using the
records of $I-V$ characteristics at 4-6 fixed values of magnetic
fields (similarly to the figure \ref{fig1}). Typical magnetic field
dependencies of the normalized saturation current $I_s(B)/I_s(0)$
for the same sample as in figure \ref{fig1} are shown in figure
\ref{fig3}. The magnitude of effect decreases with increasing
temperature (figure \ref{fig3}b) and composition $x$ (figure
\ref{fig3}a). Especially distinctly it is manifested in the range of
not large magnetic fields $B<1$ T. At higher magnetic fields the
$I_s(B)/I_s(0)$ dependencies exhibit the tendency to saturation but
do not reach overall saturation in investigated range of magnetic
fields. The $I_s(B)/I_s(0)$ values at highest available magnetic
field $B=5$ T change from diode to diode even within the series of
the samples of one type. At low temperature $T\sim 120$ K the ratio
$I_s(B_{max})/I_s(0)$ ranges from 0.04 to 0.15 and increases with
temperature.

If the above-barrier transport is due to the heavy holes, as it is
customarily supposed, we are to expect the pronounced effect only at
extremely strong magnetic fields within the framework of both diode
($B>20$T) and diffusion ($B>10$ T) theory. However experimentally
more than twofold decrease in $I_s$ is observed even at $B<0.5$ T
for sample with x=0,22 and $B<1$ T for samples with x=0,29. The
mentioned small values of $B$ correspond to $\theta\sim1$ and $\mu
B\sim1$ namely to carriers with light mass and so we can assume that
the heavy holes do not contribute significantly to the above-barrier
emission. This is in agreement with the fact that the magnitudes of
$B$ corresponding to the same value of $I_s(B)/I_s(0)$ are
approximately twice as large for material with x=0,29 than for
sample with x=0,22 (compare curves in \ref{fig3}a) in accord with
the ratio of light carriers masses and mobilities, whereas the
effective masse and mobility of heavy holes for both compositions
are the same. The contribution of heavy holes in current is
manifested by the approximate saturation of $I_s(B)$ dependencies at
high magnetic fields.

At first, we analyze $I_s(B)$ dependencies at low magnetic fields at
which the ratio $I_s(B)/I_s(0)>0.7$. In this case, as a first
approximation, one can neglect the contribution of heavy holes. In
figure \ref{fig3} the experimental $I_s(B)/I_s(0)-B$ curves are
compared with dependencies
 \begin{eqnarray}\label{Js_mobil}
 I_s(B)/I_s(0)=1/(1+(\mu_{lh}(0)B)^2),
 \end{eqnarray}
which are to be expected for effect caused by suppression of
mobility in a magnetic field. The light hole mobility in zero
magnetic field $\mu_{lh}(0)$ in (\ref{Js_mobil}) was evaluated
according to the empirical expression for electron mobility in
n-type HgCdTe \cite{mobil}
 \begin{eqnarray}\label{mobila}
 \mu_{e}=9\times10^{8}\frac{b^{7.5}}{T^{2b^{0.6}}}
 \end{eqnarray}
where $b=0.2/x$, T is in kelvin and $\mu$ in cm$^2$ V$^{-1}$
s$^{-1}$. In the low magnetic fields range the figure \ref{fig3}
shows a good fit of the relation (\ref{Js_mobil}) to the
experimental data (for heavy holes the magnitude of effect expected
at $B=5$ T does not exceed 0.3 $\%$ even for the lowest
temperatures). As the mobility decreases with temperature and
composition $x$ increase the value of effect should also decrease.
It is in agreement with experimental data presented in figure
\ref{fig3}a for samples of different composition and in figure
\ref{fig3}b for different temperatures. Though, as it is seen in
figure \ref{fig3}, the agreement (in low magnetic fields) take place
for all samples and temperatures no significance should be attached
to this fact. The point is that one can strongly overestimate the
magnitude of effect using relation (\ref{mobila}). Since p-type
material is heavily compensated, the electron mobility can be
substantially lower than the one in n-type material of similar
majority concentration \cite{Schacham}. Additionally, there are
experimental evidences that in p-type HgCdTe the light hole mobility
can be less than the electron mobility \cite{Gold}. As a result the
reduction in mobility $\mu_{lh}$ by a factor up to several times as
compare with values given by (\ref{mobila}) can be expected.

However there are two more serious physical reasons to guess that
the agreement between experimental and calculated according to
(\ref{Js_mobil}) $I_s(B)/I_s(0)$
 values in figure \ref{fig3} is most likely
casual. Firstly, the dependence of mobility on a magnetic field can
result in the suppression of a saturation current only in the case
of diffusion limited above-barrier current. At the same time the
$\mu_{lh} E_m/\overline{v}_{lh}$ ratio, usually used as a criteria
for the identification of dominant transport mechanisms in SB,
changes for investigated samples within the limits of 25$\div$2.5 in
temperature range 100$\div$250 K (very similar values are obtained
for Bete's ratio $2\varphi_{0}kT/\ell W\sim 25\div 3$; here, $\ell$
is free path and $W$ is the width of space-charge region). This
provides impressive evidence that the dominate mechanism, at least
at $T<$ 230-250 K, is the thermionic emission. It must be noted that
for heavy holes $\mu_{hh} E_m/\overline{v}_{hh}\approx 0.7\div0.2$
in the same temperature range. This fact allows to assume that for
heavy holes the diffusion current can contribute significantly.
Secondly, and it is of especial importance, the magnitude of effect
described by (\ref{Js_mobil}) is determined by a component of a
magnetic field which is perpendicular to current flow direction and
the effect should be absent in $\textbf{B}\parallel \textbf{I}$
orientation. However the measurements in tilted magnetic fields
testify that the magnitude of the effect $I_s(B)/I_s(0)$ weakly
depends on the angle between $\textbf{B}$ and $\textbf{I}$. It seen
in figure \ref{fig3} that although the effects of magnetic field for
$\textbf{B}\perp \textbf{I}$ orientation is somewhat larger than in
$\textbf{B}\parallel \textbf{I}$ orientation but the difference is
small. For low magnetic fields, at which $I_s(B)/I_s(0)>0.7$, the
magnitude of effect is practically the same in both orientations
(see the insets in figure \ref{fig3}a).

\begin{figure}[t]
\begin{center}\leavevmode
\includegraphics[scale=1.5]{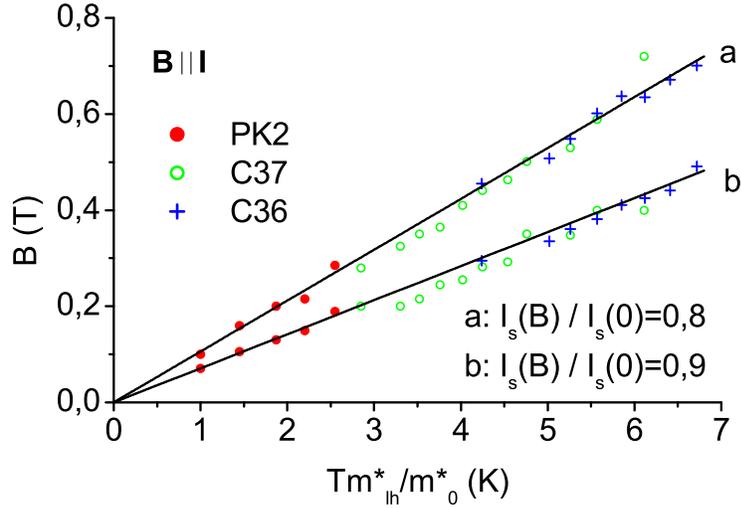}
\end{center}
\caption{Experimental dependence of magnetic field corresponding to
the same value $I_s(B)/I_s(0)$ versus the product of temperature and
light holes effective mass  with allowance for its temperature
dependence.}\label{fig4}
\end{figure}

The experimental observations discussed above indicate that the
magnetic field effect is caused by suppression of thermionic current
of light holes in magnetic field. In the case of thermionic emission
the carriers move in the space-charge region of Schottky barriers
ballistically and the current does not depend on mobility.
Meanwhile, as previously mentioned, there is strict correlation
between the magnitude of effect and the effective mass of light
carriers. The analysis of experimental data for different alloy
composition $x$ and temperatures $T\geq T_0$ ($m^{*}_{lh}$ changes
not only with $x$ but with $T$ also) shows that the values of $B$
corresponding to the same magnitude of effect at $I_s(B)/I_s(0)>0.7$
are proportional to the product $Tm^{*}_{lh}$ (this is clearly seen
in figure \ref{fig4}). From this follows the conclusion that the
value of magnetic field effect, at the least for low magnetic fields
when $I_s(B)/I_s(0)>0.7$, is uniquely determined by the ratio $
B/Tm^{*}_{lh}$ or, in energy units, by the parameter
$\theta=\hbar\omega_c/kT$.

\begin{figure}[t]
\begin{center}\leavevmode
\includegraphics[scale=1.5]{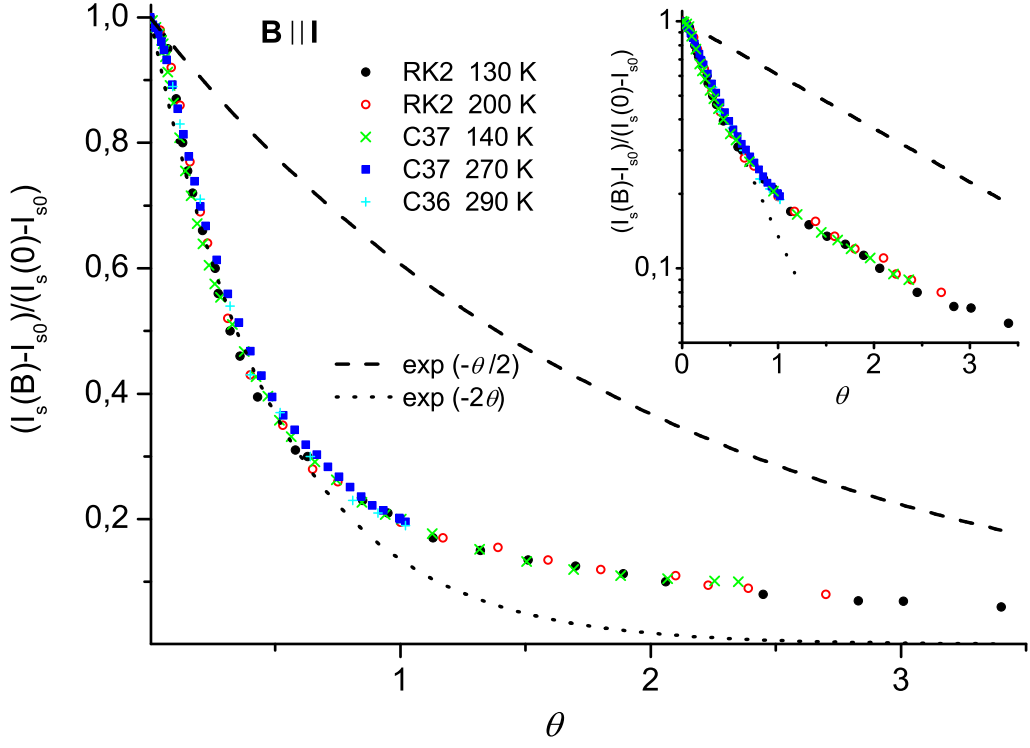}
\end{center}
\caption{Experimental dependence of fraction $(I_s(B)
-I_{s0})/(I_s(0) -I_{s0})$ versus $\theta$ parameter. The
experimental curves are matched together at a point $\theta=1$ using
$I_{s0}$ as an adjusting parameter.}\label{fig5}
\end{figure}

At higher magnetic fields the $I_s(B)$ dependencies can be affected
by the contribution of heavy holes. As mentioned above, the relative
contribution of heavy holes is different for different SB because of
uncontrolled scatter in the parameters of oxide layer (mainly the
layer thickness), and, in principle, can be temperature dependent
because of temperature dependence of light hole effective mass. To
compare $I_s(B)$ dependencies for different samples and
temperatures, and bring to light the character of $I_s(\theta)$
dependencies in high magnetic fields, the normalized saturation
current fractions $(I_s(B) -I_{s0})/(I_s(0) -I_{s0})$, controlled by
light carriers, were plotted as a function of parameter $\theta$
(see figure \ref{fig5}). Here $I_{s0}$ is a possible contribution of
heavy holes (unaffected by a magnetic field at $B<5$T) to the total
saturation current. The matching parameters  $I_{s0}$ for each
sample and the temperature were determined experimentally from the
condition $(I_s(B) -I_{s0})/(I_s(0) -I_{s0})=0.2$ at $\theta=1$.
Such relation used for matching corresponds to the SB with the
smallest saturation current at $B=5$ T. In accordance with a
difference in the values of effect at maximum magnetic field
$B\simeq 5$T the magnitudes of fraction $I_{s0}$ are different for
different samples. As a rule, the values $I_{s0}$ are larger for the
wider gap samples C36 and C37 ($J_{s0}/J_s(0)\sim 0.06\div 0.15$)
than for narrower gap sample PK2 ($I_{s0}/I_s(0)\sim 0.015\div
0.08$). $I_{s0}$ is also larger in crossed orientation as compare
with the case of $\textbf{B}\parallel \textbf{J}$. In spite of a
uncontrolled scatter in the parameters of oxide layer, the
dependence of fraction $I_s(B)-I_{s0}$ on the parameter $\theta$ for
all samples, temperature and orientation of $B$ fits common
universal curve (see figure \ref{fig5}). Thus, it is possible to
assume, that the magnitude of magnetic field effect is defined by
the ratio $\hbar\omega_c/kT $ for light holes not only at small $B$,
but in whole range of magnetic fields used here. This result along
with the absence of $B$ orientation dependence testifies that the
effect of magnetic field on the thermionic emission is due to the
magnetic quantization of light holes spectrum.

Within the framework of this mechanism the effective Kane's gap for
light carriers (gap between light holes and electron bands)
increases with magnetic field because of Landau quantization. As in
an explored interval of magnetic fields $\hbar\omega_c<<E_g$, one
can neglect the nonparabolicity. So $\Delta E_g(B)$ is given by
$\Delta E_g(B)=\hbar \omega_c=\hbar eB/m^{*}_{lh}$ (we neglect also
spin effects). On the other hand, because of a high density of
states of heavy holes band, the Fermi energy and barrier width are
determined by the majority carriers, heavy holes, while the current
is manly a contribution of light holes due to separative role of
insulator layer. Because the heavy holes band is unaffected by
magnetic field one can suppose that Schottky barrier height for
light holes $\varphi_{B}$ increases by $\Delta\varphi_{B}=n\hbar
\omega_c$, where $n=1/2$ without taking into account the spin
splitting. Within this model the dependence $I_s$ upon $B$ according
to (\ref{Ideal2a}) is expected as following:
\begin{equation}\label{I_Teta}
    \frac{I_s(B)-I_{s0}}{I_s(0)-I_{s0}}\approx\frac{I_{slh}(B)}{I_{slh}(0)}=\exp{-n\theta}.
\end{equation}

However, the experimental magnetic field dependence of saturation
current is more complicated and quantitatively does not follow this
model. The magnitude of magnetic field effect exceeds considerably
the  prediction of the simple theory (\ref{I_Teta}) (the fit by
(\ref{I_Teta}) with $n=1/2$ is shown by dashed line in figure
\ref{fig5}). For small $\theta<0.5$, the
$(I_s(B)-I_{s0})/(I_s(0)-I_{s0})$ dependence
 can be roughly described by (\ref{I_Teta}) only if the factor $n=2$ is chosen
(dotted line in figure \ref{fig5}). The reasons of this discrepancy
are not understood. It is clear, that the account for
nonparabolicity and Zeeman spliting should only decrease the
magnitude of effect expected.

In spite of the noted quantitative discrepancy there is no doubt, in
our opinion, that the transport in Schottky barriers at issue is
caused by a thermionic current of light carriers. This makes it
necessary to analyse anew the temperature dependence of a saturation
current which is usually used for determination of Richardson
constant. Such temperature dependence was discussed earlier in works
\cite{Polla,Bahir,RadFTP}.

\section {Richardson plot \label{Rich}}

\begin{figure}[t]
\begin{center}\leavevmode
\includegraphics[scale=1.7]{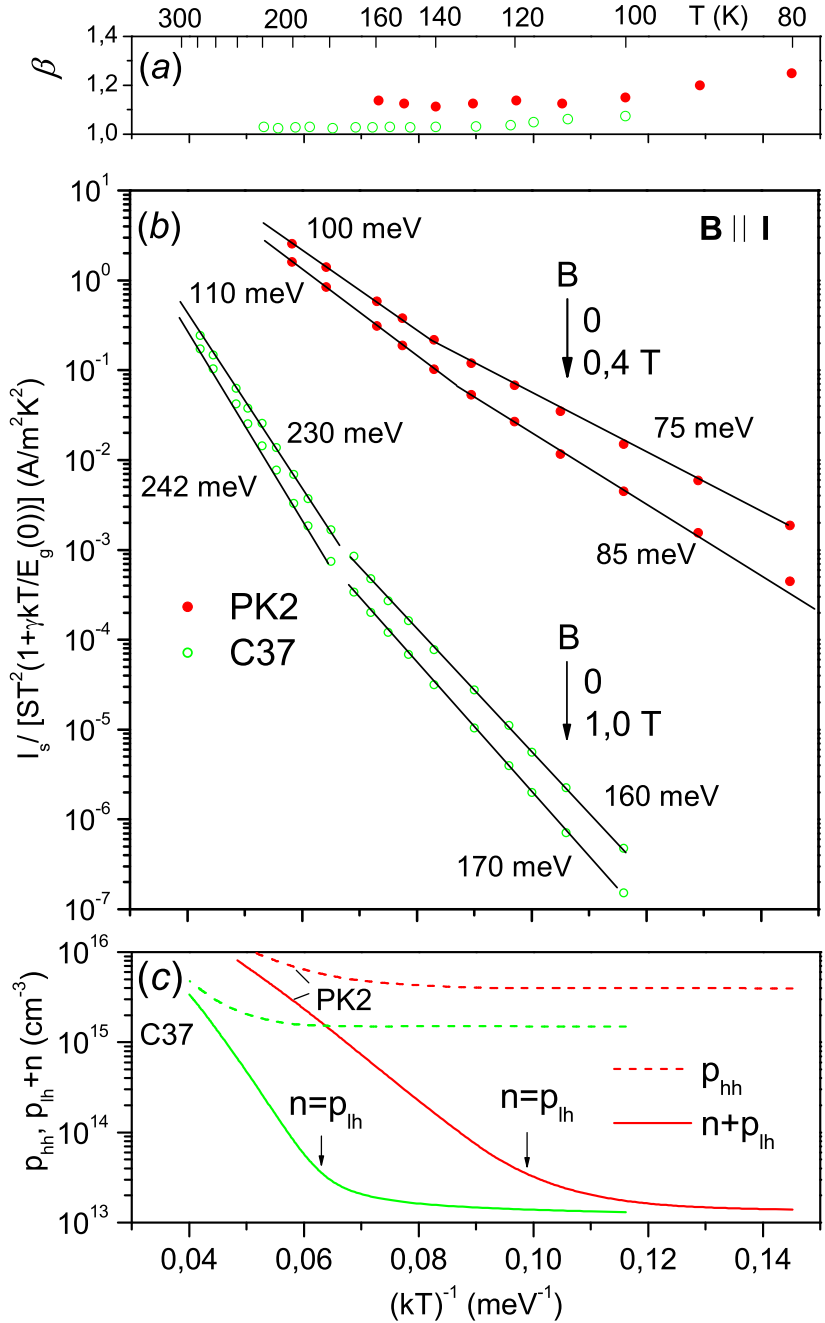}
\end{center}
\caption{Temperature dependence of the ideality factor (a), the
saturation current (b) and the concentration of carriers in
quasineutral $p-$ region of SB (c).}\label{fig6}
\end{figure}

The effective Richardson constant for heavy holes in
Hg$_{1-x}$Cd$_x$Te is expected as $A^{*}_{hh}\approx 6\times 10^5$ A
m$^{-2}$ K$^{-2}$. For light holes $A^{*}_{lh}$ is not only
composition but also temperature dependent because of temperature
dependence of light holes effective mass
$m^{*}_{lh}(T)/m^{*}_{lh}(0)=E_g(T)/E_g(0)$. For sample PK2 (C37) in
investigated temperature range  $m_{lh}$ increases almost by 80 \%
(by 30 \%) in comparison with its value at T = 0. For this reason
the standard relationship (\ref{Ideal2a}) used for determination of
Richardson constant has to be corrected
\begin{equation}\label{I_T_corr}
    \ln\frac{I_s}{ST^2\left[1+\gamma kT/E_g(0)\right]}=\ln
    A_{lh0}^{*}-\frac{\varphi_0}{kT},
\end{equation}
where $A^{*}_{lh0}$ is Richardson constant corresponding to light
holes effective mass at zero temperature. Richardson curves plotted
in this way are shown in figure \ref{fig6}. The term in square
brackets in (\ref{I_T_corr}) reduces appreciably a slope of
Richardson plot but does not affect the intercept at the ordinate.
The neglect of the temperature dependence of effective mass results
in the overestimate of barrier hight by $\sim$3-5\%.

Experimentally the linear on $T^{-1}$ dependence (\ref{I_T_corr})
retains its validity up to temperatures $T\sim 140$K for sample with
$x=0.221$ and $T\sim 180$K for samples with $x=0.29$. At higher
temperatures an increase in slope is observed. The value 150 (60) A
m$^{-2}$ K$^{-2}$ for sample RK2 (C37) is determined from the
extrapolated intercept of experimental Richardson plot
(\ref{I_T_corr}) at the ordinate  in its low-temperature linear
portion. These apparent magnitudes of Richardson constant are much
lower than their theoretical values not only for heavy holes (by
factor $\sim 5\times 10^3$), but for light holes also (by factor
$\sim 10^2$). However there are two factors, which should change the
intercept at the ordinate of experimental Richardson plot in
comparison with $A_{lh0}^{*}$: the temperature dependence of barrier
height and tunnelling through the insulator layer. When the
temperature dependence of barrier height (\ref{Fi(T)}) and
penetration coefficient (\ref{Penter}) of the interfacial insulator
layer for light holes $P_{lh}$ are taken into account, the
relationship (\ref{I_T_corr}) is modified as
 \begin{equation}\label{I_T_corr1}
    \ln\frac{I_s}{ST^2\left[1+\gamma kT/E_g(0)\right]}=\ln P_{lh} A_{lh0}^{*}-\frac{\varphi_0(0)+\alpha\gamma
    kT}{kT}.
 \end{equation}

At linear temperature dependence of energy gap $E_g$ ($\gamma
=const$) the dependence $\ln I_s/\left[1ST^2\left[1+\gamma
kT/E_g(0)\right]\right]$ - $T^{-1}$ is linear with the slope
determined by the barrier height $\varphi_0(0)$ at $T=0$. It is seen
from (\ref{I_T_corr1}) that an allowance for temperature dependence
of barrier height causes the change in a value of the intercept at
the ordinate of Richardson plot. Because in samples under
investigation $\gamma>0$ the apparent effective Richardson constant
decreases. This is contrary to the case of SB on the "classical"
semiconductors, in which Richardson constant is modified to the
higher values because of negative temperature coefficient of the
band gap and barrier height \cite{Rhoder1, Rhoder2}.

It is clear that the introduction of penetration coefficient
$P_{lh}$ is practically equivalent to the reduction of the
Richardson constant. Unfortunately, at present we have not correct
theoretical description of $I_s(B)$ dependence for light holes. If
such dependence was known, it would be possible to determine the
contribution of heavy holes by fitting the theoretical
$I_{slh}(B)+I_{shh}$ dependence to the experimental one. Using the
ratio $I_{shh}/I_{slh}$ obtained in this way one can find from the
equation (\ref{I/I}) the parameter $\delta \sqrt{\varphi_{ox}}$ and
therefore determine $P_{lh}$. As a first-order estimation for
$I_{shh}$ (upper bound), the values of a saturation current at
highest available magnetic field $B_m$ and at lowest temperature may
be used. The ratio $I_{s}(B_m)/I_s(0)$ ranges from 0.04 to 0.1.
Using these values the quantity $\delta \sqrt{\varphi_{ox}}$ can be
estimated from equation (\ref{I/I}) as $0.95-1.1$ nm eV$^{1/2}$,
which gives $P_{lh}\sim0.3 -0.35$ for $x=0.221$ and $P_{lh}\sim0.2
-0.25$ for $x=0.29$ (for heavy holes $P_{hh}\sim(4
-10)\times10^{-4}$). It should be noted that though the separative
role of oxide layer must be more pronounced for narrow-gap
semiconductors, the ratio (\ref{I/I}) weakly depends on composition
$x$ and temperature, ranging from 0.075 to 0.06 in composition
interval $x=0.17-0.4$ and temperature range $T=100-250$ K if
$\delta=1$ nm and $\varphi_{ox}=1$ eV are chosen (even for SB on
$p$-Si $I_{shh}/I_{slh}\approx 0.15$ at the same $\delta$ and
$\varphi_{ox}$). On the other hand, $I_{shh}/I_{slh}$ strongly
depends on height and thickness of a barrier. For the same
$\delta=1$ nm but for $\varphi_{ox}=1,5$ eV the equation (\ref{I/I})
gives $I_{shh}/I_{slh}=0.019-0.017$ in the same temperature and
composition ranges.

As may be inferred from (\ref{I_T_corr1}) the true Richardson
constant $A_{lh0}^{*}$ and apparent effective Richardson constant
$A_{lh0}^{**}$ (experimentally determined from Richardson plot as
the intercept at the ordinate) are related by the expression
 \begin{equation}\label{AA}
  A_{lh0}^{*}=\exp(\alpha\gamma/k)P_{lh}^{-1}A_{lh0}^{**}.
 \end{equation}
The Richardson constant $A_{lh0}^{*}$ for sample RK2 is calculated
from this relation as $8.2\times10^3$ A m$^{-2}$ K$^{-2}$  taking
the above determined values of $A_{lh0}^{**}$ and $P_{lh}$. This is
in close agreement with the theoretically expected value of
$8.8\times10^3$ A m$^{-2}$ K$^{-2}$. The similar calculation for
samples C37 and C36 gives $A_{lh0}^{*}\sim (2.2-2.5)\times10^3$ A
m$^{-2}$ K$^{-2}$ that is less than expected value of $2\times10^4$
A m$^{-2}$ K$^{-2}$.

In figure \ref{fig6} the Richardson curves measured in low magnetic
fields at which the heavy holes do not contribute significantly are
also plotted. As one can see no peculiarities in Richardson plot in
magnetic field as compare with $B=0$ case are observed. The effect
of magnetic field is manifested only as an increase in the slope of
Richardson plots that indicates the rise of Schottky barrier height.
The apparent effective Richardson constant $A_{lh0}^{**}$ is
practically unaffected by a magnetic field. The slopes of Richardson
plots increase by $\Delta \varphi_b\sim 10-12$ meV in magnetic
fields used. These values are approximately 3-4 times higher than
those expected for $\Delta\varphi_{B}=\hbar \omega_c/2$ for light
holes ($\Delta\varphi_{B}\approx 2.3$ meV for $x=0.221$ and
$\Delta\varphi_{B}\approx 3$ meV for $x=0.29$). Thus, there is the
same inconsistency, that was already noticed in section \ref{IB} at
the analysis of magnetic field dependencies of saturation current.

As already  mentioned, the experimental $I_s(T)$ dependencies can
not be described as a whole by the equation (\ref{I_T_corr1}) in all
temperature range. The Richardson plot is in fact composed of two
linear regions (figure \ref{fig6}b). The previous analysis in this
section was concerned with a low-temperature portion of Richardson
plot. At higher temperatures the slope for both samples increases to
the value close to $E_g$. The intercept at the ordinate for this
region for sample PK2 (C37 and C36) is order (two order) of
magnitude higher than its value for low-temperature portion. These
values are of order of magnitude higher than the theoretically
expected value of Richardson constant $A_{lh0}^{*}$ for light
carriers.

The possible reason for "two-region" behaviour of the $\ln I_s(T)$
dependencies is contribution of minority carriers (electrons) to
total current at high temperatures. Because heavy holes are removed
out of the transport by the interface layer, the contribution of
electrons to current may be significant already at moderate
temperatures at which the transition to intrinsic conductivity is
only beginning. In figure \ref{fig6}c the temperature dependence of
$p_{hh}$  and $p_{lh}+n_e$ calculated from the electroneutrality
equation while taking into account the band nonparabolicity for
light carriers and the temperature dependence of Kane's gap $E_g$
and effective mass $m^{*}_{lh}$ and $m^{*}_{e}$ are plotted. It is
seen that the noticeable increase in $p_{hh}$ is beginning at
$T\sim180$ K ($x=0.22$) and $T\sim220$ K ($x=0.29$) whereas the
contribution of electrons in total concentration of light carriers
is essential already at $T\sim120$ K ($x=0.22$) and $T\sim160$ K
($x=0.29$) (the temperatures at which $n_e=p_{lh}$ are marked by
arrow in figure \ref{fig6}c). Namely above these temperatures the
change in slope of Richardson plot is observed. The activation
energy of $p_{lh}+n_e$ curves for this transition region of $T$ is
close to $E_g$ (at higher $T$ this energy approaches to $E_g/2$)
what is close to the slope of high-temperature region of Richardson
plot. Figure \ref{fig6}b shows that in magnetic field an increase in
the slope of Richardson plot in its high-temperature portion is
within the experimental error the same as in low-temperature range
in agreement with equality of the cyclotron energies of light holes
and electrons.

\section{Conclusion}

The strong magnetic field effect on above-barrier transport is
revealed and investigated in wide temperature interval in Schottky
barriers on $p$- type narrow-gap Hg$_{1-x}$Cd$_x$Te. The large
magnitude of effect and its weak dependence on magnetic field
orientation indicates that the dominant transport mechanism is
thermionic emission of light holes confirming the conclusion based
on Growell-Zee and Bete criteria. The predominance of light holes in
ballistic transport in SB based on $p$ type semiconductors should be
expected because of low tunnel penetrability of oxide layer at
interface for heavy holes. This assumption together with temperature
dependence of barrier height satisfactorily explains the
experimental values of Richardson constant.

The investigations performed for materials with different band
parameters and at different temperatures indicate that the magnitude
of magnetic field effect is uniquely determined by the ratio of
light hole cyclotron energy to a thermal energy
$\theta=\hbar\omega_{clh}/kT$. However there is an essential
discrepancy between the experimental and predicted dependencies of a
saturation current on a $\theta$ parameter. To throw light on this
discrepancy it would be interesting to investigate the magnetic
field effect in similar SB based on $n-$ type narrow gap
Hg$_{1-x}$Cd$_x$Te. Unfortunately, as earlier it was marked, such
barriers do not exhibit the rectifying properties. Another possible
subject of inquiry is  magnetic field effect on the diffusion
transport in Hg$_{1-x}$Cd$_x$Te $p-n$ junctions. The first
preliminary results of such research we performed reveal the strong
effect at both orientations of a magnetic field. The magnetic-field
dependence of saturation current in $p-n$ junction is found to be in
general similar to that for SB, but there are some specific features
needing special consideration that is beyond the scope of this
paper.

Due to the separative role of oxide layer the light holes
predominate not only in the tunnelling current \cite{RadFTP} but in
the thermionic current also. As a result there is no replace of
carriers responsible for transport in SB on $p-$ type HgCdTe at
transition from a regime of tunnelling at low temperatures $T<T_0$
to a regime of thermionic emission at $T>T_0$. In both cases the
light holes carry over a current. It should be noted that the
discussed effect of insulator layer on thermionic current does not
concern the results of works \cite{RadFTP,RadSST} concidering the
tunnelling and magneto-tunnelling in SB on $p-$ HgCdTe at low
temperatures $T<T_0$. The influence of this layer on the tunnelling
currents is negligibly small because its tunnelling transparency for
light holes $P_{lh}\sim0.2-0.4$ many orders over exceeds the
transparency of depletion layer $P_{SB}\sim\exp(-\lambda_0)$ where
$\lambda_0=5-30$ \cite{RadSST}.

The prevalence of light holes in thermionic current should be taken
into account at the analysis of transport in Schottky barriers based
on the  $p$- type semiconductors, especially in
metal-insulator-semiconductor Schottky diodes \cite{Si_CRT_InP,
MIS_Si,MIS_Hg}.


\section*{References}
\begin{thebibliography}{99}
\bibitem{Sze} Sze S and Ng K 2008 {\it Physics of Semiconductor Devices} 3nd edn
(New York: Wiley)
\bibitem{Rhoder} Rhoderick E H and Williams R H 1988 {\it Metal–Semiconductor Contacts} 2nd edn
(Oxford: Claredon)
\bibitem{Tun_mass} Williams G M and DeWames R E 1995 {\it J. Electron. Mater.} {\bf 24} 1239
\bibitem{Polla} Polla D L and Sood A K 1980 \JAP {\bf 51} 4908
\bibitem{Bahir} Bahir G, Adar R and Fastov R 1991 {\it J. Vac. Sci. Technol.} A {\bf
9} 266
\bibitem{RadFTP} Zav`yalov V V, Radantsev V F and Deryabina T I 1992 {\it Fiz. Tekn. Poluprov.} {\bf 26}
691 (Engl. Transl. 1992 {\it Sov. Phys. - Semicond.} {\bf 26} 388)
\bibitem{RadSST} Zav`yalov V V and Radantsev V F 1994 \SST {\bf 9} 281
\bibitem{Eg(T)} Seller D G and Lowney J R 1990 {\it J. Vac. Sci.
Technol.} A {\bf 8} 1237
\bibitem{mobil} Rogalsky A 2005 \RPP {\bf 68} 2267
\bibitem{Schacham} Schacham S E and Finkman E 1988 {\it J. Vac. Sci. Technol.} A {\bf 7}
367
\bibitem{Gold} Gold M C and Nelson D A 1986 {\it J. Vac. Sci. Technol.} A {\bf
4} 2040
\bibitem{Rhoder1} Missous M and Rhoderick E H 1991 \JAP {\bf 69}
7142
\bibitem{Rhoder2} Missous M, Rhoderick E H, Woolf D A and Wilkes S P 1992 \SST {\bf 7}
218
\bibitem{Si_CRT_InP} Mikhelashvili V, Eisenstein G, Garber V, Fainleib F, Bahir G,
Ritter D, Orenstein M and Peer A 1999 \JAP {\bf 85} 6873
\bibitem{MIS_Si} Yildiz D E, Altindal S, and H. Kanbur H 2008 \JAP {\bf 103} 124502
\bibitem{MIS_Hg} Damnjanovi´c Vesna, Ponomarenko V P  and Elazar Jovan M 2007 \SST {\bf 22}
137

\end{thebibliography}
\end{document}